\documentclass[aps,pra,showpacs,twocolumn,superscriptaddress,groupedaddress]{revtex4-1}

\usepackage{graphicx}% Include figure files
\usepackage{amsmath,amssymb}
\usepackage{color}% Include color
\usepackage{comment}% Include comment

\begin{document}

\title{
Magnetism in the three-dimensional layered Lieb lattice: \\
 Enhanced transition temperature via flat-band and Van Hove singularities
}

\author{Kazuto Noda}
\email{noda.kazuto@lab.ntt.co.jp}
\affiliation{NTT Basic Research Laboratories, NTT Corporation, Atsugi 243-0198, Japan}
\author{Kensuke Inaba}
\affiliation{NTT Basic Research Laboratories, NTT Corporation, Atsugi 243-0198, Japan}
\author{Makoto Yamashita}
\affiliation{NTT Basic Research Laboratories, NTT Corporation, Atsugi 243-0198, Japan}

\begin{abstract}
We describe the enhanced magnetic transition temperatures $T_c$ of two-component fermions in three-dimensional layered Lieb lattices, which are created in cold atom experiments.
We determine the phase diagram at half-filling using the dynamical mean-field theory.
The dominant mechanism of enhanced $T_c$ gradually changes from the (delta-functional) flat-band  to the (logarithmic) Van Hove singularity as the interlayer hopping increases.
We elucidate that the interaction induces an effective flat-band singularity from a dispersive flat (or narrow) band.
We offer a general analytical framework for investigating the singularity effects, where a singularity is treated as one parameter in the density of states.
This framework provides a unified description of the singularity-induced phase transitions, such as magnetism and superconductivity, where the weight of the singularity characterizes physical quantities.
This treatment of the flat-band provides the transition temperature and magnetization as a universal form (i.e., including the Lambert function).
We also elucidate a specific feature of the magnetic crossover in magnetization at finite temperatures.
\end{abstract}

% pacs
% 05.30.Fk   Fermion systems and electron gas (see also 71.10.-w Theories and models of many-electron systems; see also 67.10.Db Fermion degeneracy in quantum fluids)
% 37.10.Jk   Atoms in optical lattices
% 67.85.-d Ultracold gases, trapped gases (see also 03.75.-b Matter waves in quantum mechanics)
% 71.10.Fd   Lattice fermion models (Hubbard model, etc.)
% 71.27.+a Strongly correlated electron systems; heavy fermions
% 75.10.-b General theory and models of magnetic ordering (see also 05.50.+q Lattice theory and statistics)
\maketitle

%%%%%%%%%%%%%%%%%%%%%%%%%%%%%
\section{Introduction}
%%%%%%%%%%%%%%%%%%%%%%%%%%%%%

Phase transitions, such as magnetism and superconductivity, are of fundamental interest in lattice fermions. 
As a common feature, transition temperatures $T_c$ are usually given as a function of the density of states (DOS) at the Fermi energy $\rho(E_F)$ and an interaction $U$, $T_c/W\propto e^{-1/\rho(E_F)U}$ for $U/W\ll 1$, where bandwidth $W$ is a unit of energy.
A singularity located on the Fermi energy [$\rho(E_F)\to \infty$] changes this functional form, which could greatly increase the $T_c$.
For instance, the logarithmic Van Hove singularity (VHS) induces a characteristic dependence: $T_c/W\propto e^{-\sqrt{W/U}}$\,\cite{hirsch_enhanced_1986}.
Recent studies proposed the emergence of another interesting delta-functional singularity, which we call the flat-band singularity (FBS), at the surface of a layered graphene \cite{Kopnin2011} or of a topological material \cite{Tang2014}. 
This FBS is also expected as the origin of the higher transition temperature: $T_c\propto U$.
Although these singularities have attracted attention, we still lack a comprehensive understanding of these singularity effects on phase transitions.

Cold atoms in an optical lattice, where we can control lattice geometry and resulting DOS singularity \cite{Bloch2008,windpassinger_engineering_2013}, provide opportunities for studying the singularity effects in bulk systems.
In particular, successful creations of two-dimensional (2D) optical lattices with singular DOSs, the Kagome \cite{Jo2012} and Lieb (line-centered-square) \cite{takahashi} lattices, have activated theoretical studies on phenomena related to the FBS
\cite{Lieb1989,Mielke1992,Noda2009,Shen2010,Apaja2010,Weeks2010,Huber2010,Goldman2011,Zhao2012,You2012,Yamamoto2013,Iglovikov2014}.
In general, in these 2D lattices, the Mermin-Wagner theorem states that no phase transitions occur at finite temperatures \cite{Mermin1966}.
A layered structure exhibits specific features in the DOS as a remnant of 2D lattices, even though the system itself is three-dimensional (3D) \cite{Noda2014}.
Here we focus on the 3D layered Lieb lattice \cite{takahashi,Noda2014}, which is a test-bed for systematic investigations of the effects of various singularities on phase transitions.

In this paper, we show that the magnetic transition temperature $T_c$ is clearly enhanced by the FBS and VHS in a 3D layered Lieb lattice at half-filling using the dynamical mean-field theory (DMFT).
We determine the phase diagrams with several anisotropic hoppings between the inter- and intralayer directions.
It is shown that this anisotropy is a practical parameter for controlling $T_c/W$ for both weakly and strongly interacting regions.
We also propose an analytical framework for dealing with the singularity effects with a single parameter called a singularity weight $a$.
We demonstrate that $T_c\propto aU$ for FBS and $T_c/W \propto e^{-\sqrt{2W/aU}}$ for VHS.
These forms clearly provide a unified picture of phase transitions dominated by singular DOSs, where a large singularity-weight enhances $T_c$.
This is in stark contrast to phase transitions in nonsingular systems.
We also demonstrate that, for both singularities, a conventional form $e^{-1/\rho(E_F)U}$ can universally be reproduced for $a\to 0$.
Our approach also phenomenologically explains that the FBS induces an anomalous behavior of thermodynamic quantities even in the paramagnetic state above $T_c$.

%%%%%%%%%%%%%%%%%%%%%%%%%%%%%
\section{Model and Method}
%%%%%%%%%%%%%%%%%%%%%%%%%%%%%
\label{sec_modelandmethod}

\begin{figure}[tb]
 \begin{minipage}[c]{0.43\linewidth}
  \begin{center}
   \includegraphics[clip,width=\linewidth]{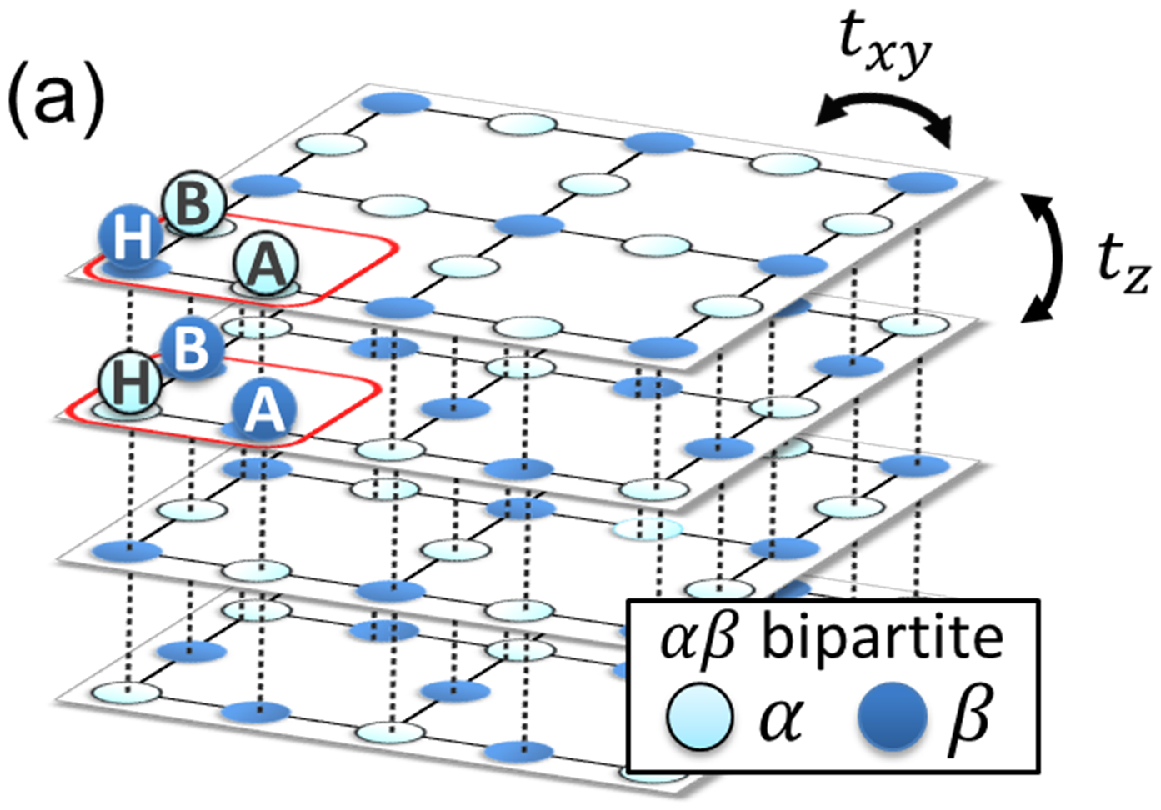}
  \end{center}
 \end{minipage}
 \begin{minipage}{0.48\linewidth}
  \begin{center}
  \includegraphics[clip,width=\linewidth]{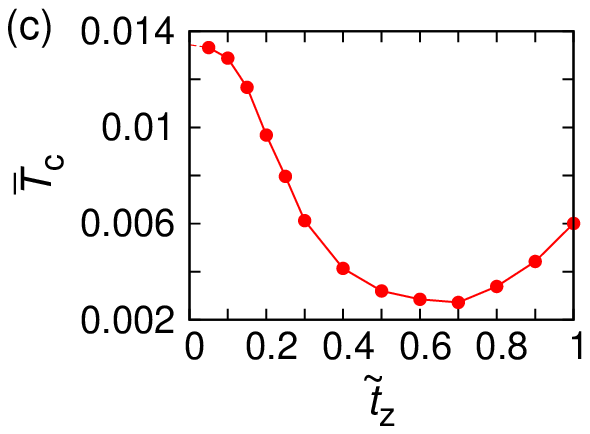}
  \end{center}
 \end{minipage}
\begin{minipage}{0.7\linewidth}
  \begin{center}
   \includegraphics[clip,width=\linewidth]{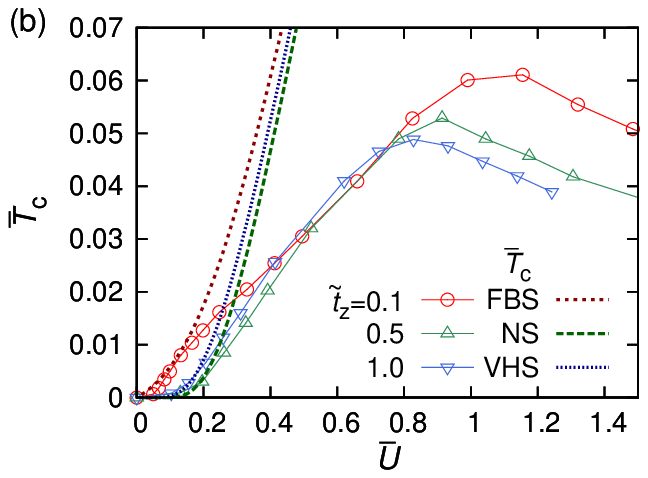}
  \end{center}
 \end{minipage}
\caption{(Color online) (a) 3D layered Lieb lattice. Solid lines represent the unit cell for our calculations. 
(b) Transition temperature $\bar{T}_c$ vs. interaction $\bar{U}$ for $\tilde{t}_z=0.1, 0.5$, and $1.0$.
The thick dotted lines are analytical forms: 
$\bar{T}_c^{\text{FBS}}\propto a\bar{U}$,
$\bar{T}_c^{\text{NS}}\propto e^{-1/\bar{U}}$ and
$\bar{T}_c^{\text{VHS}}\propto e^{-1/\sqrt{a\bar{U}}}$.
See text for their full forms.
(c) $\bar{T}_c$ vs. $\tilde{t}_z$ for $\bar{U}=0.2$.
}
\label{fig_tn}
\end{figure}

Our model is described by the Hubbard Hamiltonian on a 3D layered Lieb lattice [see Fig.\,\ref{fig_tn}\,(a)]:
\begin{equation}
\nonumber
%\label{eq_3dlieb}
{\cal H}
=
- t_{xy} \sum_{l\langle i,j \rangle \sigma} c_{li\sigma}^{\dagger} c_{lj\sigma}
- t_z \sum_{\langle l,l' \rangle i \sigma} c_{li\sigma}^{\dagger} c_{l'i\sigma}
+ U \sum_{li} n_{li\uparrow}n_{li\downarrow},
\end{equation}
where $c_{li\sigma} (c_{li\sigma}^{\dagger})$ is the annihilation (creation) operator of an atom with spin $\sigma$ at site $i$ on the $l$-th layer, and $n_{li\sigma}=c_{li\sigma}^{\dagger}c_{li\sigma}$.
The subscript $\langle i,j \rangle (\langle l,l' \rangle)$ denotes the summation over the nearest neighbor sites in the $xy$ plane ($z$ direction). We impose periodic boundary conditions for all $x, y, z$ directions.

To investigate the magnetic properties of this model at half-filling, we use the DMFT approach \cite{Georges1996,Noda2014} with a six-site unit cell as shown in Fig.\,\ref{fig_tn}\,(a).
The bipartite structure allows us to focus on the antiferromagnetic ordering.
We employ the numerical renormalization group method (NRG) \cite{Wilson1975,Bulla2008} to solve the effective impurity problem.
NRG is applicable to energy scales ranging from the ground state to finite temperatures \cite{Anders2005,Peters2006,Weichselbaum2007}. 
Our approach (DMFT+NRG) succeeded in studying the ground state properties of the present system \cite{Noda2014}, suggesting the validity of the application to study finite temperature properties. The numerical procedures are detailed in Ref.\,\cite{Noda2014}.

Here, we choose bandwidth $W(=4\sqrt{2}t_{xy}+4t_z)$ as the unit of energy and use the notation $\bar{X}\equiv X/W$ to describe dimensionless parameters.
As an exception, we use $\tilde{t}_z\equiv t_z/t_{xy}$ to characterize the anisotropy of intralayer ($xy$-plane) and interlayer ($z$-direction) hoppings.
The DOS $\rho(\omega)$ is rescaled as $\bar{\rho}(\bar{\omega})\equiv W \rho(\bar{\omega})$.
We calculate thermodynamic quantities, {i.e.}, magnetization $m_{\alpha}=(n_{\alpha \uparrow}-n_{\alpha\downarrow})/2$ and double occupancy $D_{\alpha}=\langle n_{\alpha \uparrow}n_{\alpha\downarrow}\rangle$, where $\alpha=H, A, B$.
Note that the lattice geometry and the half-filling condition result in symmetry e.g. $m_A=m_B$ and $D_A=D_B$, and so on.

%%%%%%%%%%%%%%%%%%%%%%%%%%
%%%%%%%%%%%%%%%%%%%%%%%%%%%%

%%%%%%%%%%%%%%%%%%%%%%%%%%%%%
\section{Phase diagram}
\label{sec_phasediagram}
%{\it Phase diagram}---
We first overview the characteristic magnetism of the present system based on the phase diagram.
Figure\,\ref{fig_tn}\,(b) shows $\bar{T}_c$ as a function of $\bar{U}$ for $\tilde{t}_z=0.1, 0.5$, and $1.0$.
Below $\bar{T}_c$, the antiferromagnetic insulating states appear, while above $\bar{T}_c$, non-magnetic (metallic or Mott insulating) states appear.
From nonmonotonic $\bar{T}_c$ curves, we can see that crossovers from the band picture to the Heisenberg (local) picture of magnetic transitions occur at around $\bar{U} \sim 0.8$-$1.0$ for all $\tilde{t}_z$.

We next discuss how anisotropy $\tilde{t}_z$ affects $\bar{T}_c$.
For clarity, we provide a change in $\bar{T}_c$ in a weakly interacting region $\bar{U}=0.2$ in Fig.\,\ref{fig_tn}\,(c).
For $\bar{U}\lesssim 1$, $\bar{T}_c$ is strongly enhanced for small $\tilde{t}_z(=0.1)$.
Surprisingly, $\bar{T}_c$ for $\tilde{t}_z=0.1$ shows specific behavior $\bar{T}_c\propto \bar{U}$, which is 
qualitatively distinct from the well-known conventional weak-interacting behavior $\bar{T}_c\propto \exp(-1/\bar{U})$ (as found in those for $\tilde{t}_z=0.5$).
We find that $\bar{T}_c$ also slightly increases for $\tilde{t}_z\to 1$, which is the result of another distinct behavior of $\bar{T}_c\propto \exp(-1/\sqrt{\bar{U}})$
\cite{hirsch_enhanced_1986}.
These qualitative changes in the magnetism will be discussed in detail below with an analytical approach [see thick dotted lines in Fig.\,\ref{fig_tn}\,(b)].

For the strongly interacting region $\bar{U}\gtrsim 1$ in Fig.\,\ref{fig_tn}\,(b), we find that $\bar{T}_c$ is enhanced as $\tilde{t}_z$ decreases. 
In contrast to the above, this enhancement can be understood quantitatively: $\bar{T}_c$ can be scaled by the effective Heisenberg parameter $\bar{J}_{\rm Hei}$. 
Interestingly, in this localized-spin picture region, the difference in the number of adjacent sites plays an important role, as discussed in Sec.\,\ref{sec_stronglyinteracting}.

%%%%%%%%%%%%%%%%%%%%%%%%%%
\begin{figure}[tb]
 \begin{minipage}{0.98\linewidth}
  \begin{center}\
   \includegraphics[clip,width=\linewidth]{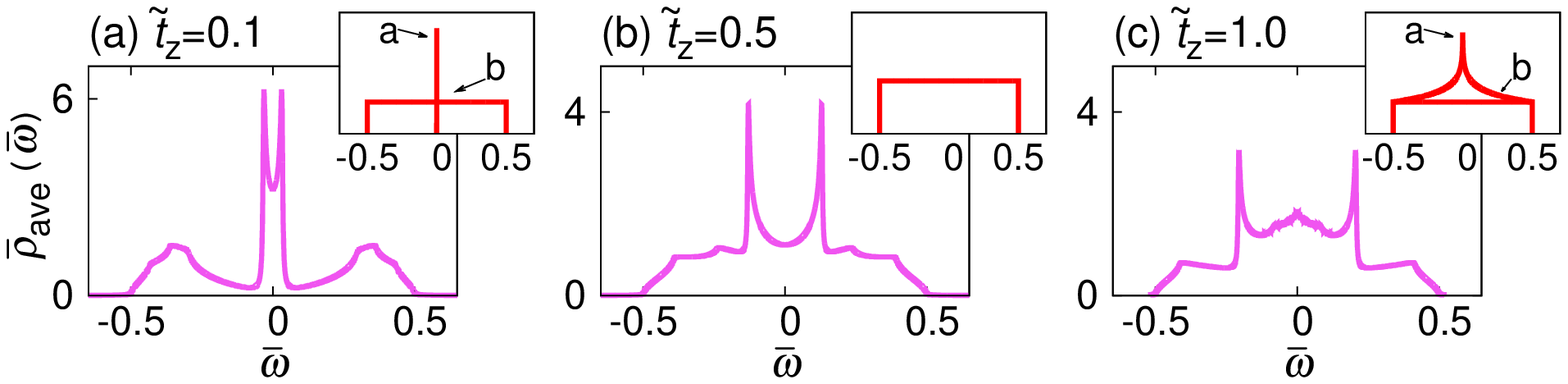}
  \end{center}
 \end{minipage}
 \begin{minipage}{\linewidth}
  \begin{center}
   \includegraphics[clip,width=0.65\linewidth]{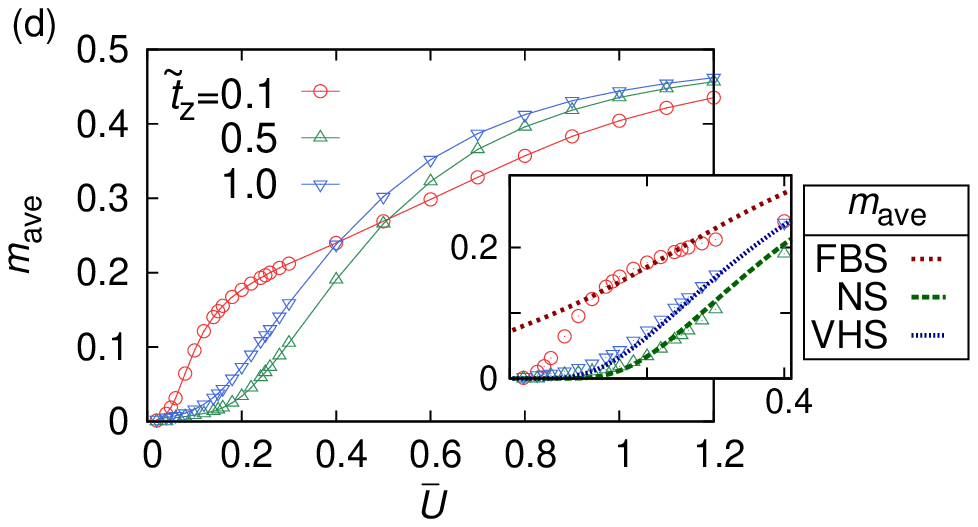}
  \end{center}
 \end{minipage}
\begin{center}
\caption{(Color online) (a)-(c) Average noninteracting DOSs for $\tilde{t}_z=0.1, 0.5, 1.0$ and (d) average magnetizations $m^{\text{ave}}$ vs. $\bar{U}$ for $\tilde{t}_z=0.1, 0.5$, and $1.0$ at zero temperature.
Insets of (a)-(c) represent the simple DOSs obtained by extracting essential structures of $\rho_{\rm ave}(\omega\sim 0)$ with normalized condition $a+b=1$: 
$\bar{\rho}^{\text{FBS}}(\bar{\omega})=a\delta(\bar{\omega})+b \bar{\rho}^{\text{uni}}(\bar{\omega})$, $\bar{\rho}^{\text{uni}}(\bar{\omega})=\theta(1/2-|\bar{\omega}|)$, and $\bar{\rho}^{\text{VHS}}(\bar{\omega})=a\ln(1/2|\bar{\omega}|)\theta(1/2-|\bar{\omega}|)+b \bar{\rho}^{\text{uni}}(\bar{\omega})$, respectively.
The inset of (d) is a closeup around $\bar{U}=0$ of $m^{\text{ave}}$ for all three $\tilde{t}_z$ with analytical results (thick dotted lines):
$m^{\text{FBS}}_{\rm ave}\sim a/2+(ab\bar{U}/2)\ln(2/a\bar{U})$ with $a=0.14$,
$m^{\text{NS}}_{\rm ave}= e^{-1/\bar{U}}/\bar{U}$,
and $m^{\text{VHS}}_{\rm ave}\propto e^{-\sqrt{2/a\bar{U}}}/\bar{U}$ with $a=1/3\pi$.
See text for the full form of $m^{\text{FBS}}_{\rm ave}$ and $m^{\text{VHS}}_{\rm ave}$.
}
\label{fig_dos}
\end{center}
\end{figure}
%%%%%%%%%%%%%%%%%%%%%%%%%%%%s

%%%%%%%%%%%%%%%%%%%%%%%%%%%%%
\section{Weakly interacting region}
\label{sec_weaklyinteracting}
%{\it{Weakly interacting region}}---
The magnetism in this region will be characterized by the band structure, and, in particular, the DOS at Fermi energy $\omega=0$ is an important quantity.
In Fig.\,\ref{fig_dos}\,(a)-(c), we thus provide 
$\bar{\rho}_{\rm ave}(\bar{\omega})[=\sum_{\alpha=H,A,B}\bar{\rho}_{\alpha}(\bar{\omega})/3]$ for $\bar{U}=0$.
As is well known, the 2D Lieb lattice ($\tilde{t}_z=0$) has a flat band \cite{Noda2009}, and therefore DOS has the delta function singularity at the Fermi energy (not shown).
For the present systems with a finite $\tilde{t}_z (\not=0)$, the flat band becomes dispersive with a width of $4t_z$.
From Fig.\,\ref{fig_dos}\,(a), we can see that a large DOS still appears for a small $\tilde{t}_z$.
This remaining flat band structure is naively regarded as the origin of the specific behavior $\bar{T}_c\propto \bar{U}$.
As shown in Fig.\,\ref{fig_dos}\,(b), the DOS at $\bar{\omega}=0$ decreases with increasing $\tilde{t}_z$, and then the FBS disappears, resulting in the conventional $\bar{T}_c\propto e^{-1/\bar{U}}$ for $\tilde{t}_z \sim 0.5$.
Figure\,\ref{fig_dos}\,(c) shows another interesting feature of the DOS for $\tilde{t}_z=1$: a logarithmic singularity at $\bar{\omega}=0$ like a 2D VHS, leading to $\bar{T}_c\propto e^{-1/\sqrt{\bar{U}}}$ \cite{hirsch_enhanced_1986}.

These results suggest that, generally, a singularity of DOS changes the functional forms of $T_c$ and then drastically enhances $T_c$.
Some previous studies have discussed logarithmic VHS effects in the 2D square
lattice \cite{hirsch_enhanced_1986,Hirsch1985} and also investigated the FBS effects \cite{Imada2000,Kopnin2011}.
However, to the best of our knowledge, a general analytical formalism for dealing with singularity effects has not yet been established.

Here, we propose a general approach for revealing the singularity effects, which can explain the $\bar{U}$ dependence of $\bar{T}_c$ in Fig.\,\ref{fig_tn}\,(b).
We consider the mean-field gap equation
\begin{equation}
\frac{1}{U}=\int d\omega  \frac{\rho_{\rm ave}(\omega)}{2\sqrt{\omega^2+\Delta_{\rm ave}^2}}\tanh\bigg(\frac{\sqrt{\omega^2+\Delta_{\rm ave}^2}}{2T}\bigg), \label{eq_gap}
\end{equation}
where $\rho_{\rm ave}(\omega)$ and $\Delta_{\rm ave}$ are the average DOS and spectral gap, respectively, with respect to sites $A, B$, and $H$.
Gap $\Delta_{\rm ave}$ can be rewritten as $\Delta_{\rm ave}=U m_{\rm ave}$, where the average magnetization $m_{\rm ave}=\sum_{\alpha= A,B, H}|m_\alpha|/3$.
We simplify the multiband structure and the specific lattice structure, which leads to the above site-averaged gap equation.
The average DOS can be set to a simple sum of the singular and the nonsingular parts: $\bar{\rho}_{\rm ave}(\bar{\omega})= a \bar{\rho}^{\rm S}(\bar{\omega})+ b \bar{\rho}^{\rm NS}(\bar{\omega})$ [see insets of Fig.\,\ref{fig_dos}\,(a)-(c)].
Here, we introduce the specific parameter $a$ ($b$) defined as a weight of the singular (nonsingular) DOS with normalization condition $a+b=1$.
We simply set $\bar{\rho}^{\rm NS}$ as the uniform DOS $\bar{\rho}^{\rm uni}(\bar{\omega})=\theta(1/2-|\bar{\omega}|)$, where $\theta(\omega)$ is a step function.

Here, we should comment that our approach provides a general extension of the conventional forms of $T_c$ and $\Delta$ [$\propto e^{-1/\rho(E_F)U}$].
A divergent $\rho(E_F)$ cannot parameterize $T_c$ and $\Delta$ any more, and instead of this, the singularity weight $a$ determines these physical quantities.
Note that, generally, any singularities of $\rho(\omega)$ should disappear in an integral and a weight $a$ is always definable: Namely, $\int \rho(\omega) d\omega (\equiv a+b) =1$ even though ${}^\exists \omega \in \mathbb{R}; \rho(\omega)=\infty$.
Thus, our approach is applicable to any singularities on any lattice geometry.

In what follows, we show that the above simplified approach with a parameterized singularity can capture the essence of the magnetic transition for $\bar{U}\ll 1$.
We start with a general discussion with $a$ of any value, which qualitatively explains the $\bar{U}$ dependence of $\bar{T}_c$. After that, with a given $a$, we quantitatively compare the analytical form with the numerical results.
We also show that the introduction of singularity weight $a$ allows us to phenomenologically understand the anomalous behavior of some thermodynamic quantities.

We first discuss the linear-$\bar{U}$ behavior of $\bar{T}_c$ shown in Fig.\,\ref{fig_tn}(b). %induced by the FBS.
We here consider a DOS with the FBS given by $\bar{\rho}^{\rm FBS}_{\rm ave}(\bar{\omega})= a \delta(\bar{\omega})+ b \bar{\rho}^{\rm uni}(\bar{\omega})$ shown in the inset of Fig.\,\ref{fig_dos}\,(a).
By solving Eq.\,(\ref{eq_gap}) with $\Delta_{\rm ave}=0$, we obtain the transition temperature (see Appendix \ref{sec_a_fbs}):
\begin{equation}
\label{eq_tn_FBS}
\bar{T}_c^{\rm FBS} =
a \left/
{4b \mathcal{W}\left(\frac{a\pi}{4be^{\gamma}} e^{1/b\bar{U}}\right)},
\right.
%^{-1},
%(=L(U/W)),
\end{equation}
where $\gamma$ is the Euler constant and $\mathcal{W}(x)$ is the Lambert function defined as $x=\mathcal{W}(x)e^{\mathcal{W}(x)}$.
For $a=0$, given $\mathcal{W}(x)\sim x$ for $x \sim 0$, Eq.\,(\ref{eq_tn_FBS}) reproduces the conventional behavior without the singularity: $\bar{T}^{\rm NS}_c=e^{\gamma-1/\bar{U}}/\pi$.
For $\bar{U}\to 0$, except for $a=0$, the divergent argument $e^{1/b\bar{U}}$ requires another asymptotic property $\mathcal{W}(x)\sim \ln x-\ln(\ln x)$ for $x \rightarrow +\infty$.
We thus obtain $\bar{T}^{\rm FBS}_c\sim a\bar{U}/4+(ab\bar{U}^2/4)\ln(4e^\gamma/a\pi\bar{U})$ for $\bar{U}\sim 0$.
This explains $\bar{T}_c\propto a\bar{U}$ shown in Fig.\,\ref{fig_tn}\,(b) and the strong enhancement of $\bar{T}_c$ in Fig.\,\ref{fig_tn}\,(c).
Importantly, this asymptotic behavior with a divergent term indicates that, even if a singularity weight $a$ is very small, the FBS changes the nature of the transition at around $\bar{U}\sim 0$
\footnote{These results can be applicable to the high-temperature surface superconductivity induced by the partially flat-band, which are discussed in Ref.\,\cite{Kopnin2011,Tang2014}.}.

We next describe the enhancement of $\bar{T}_c$ due to the VHS.
Here we consider the  DOS $\bar{\rho}^{\rm VHS}_{\rm ave}(\bar{\omega})= a \ln(1/2|\bar{\omega}|)\theta(1/2-|\bar{\omega}|)+ b \bar{\rho}^{\rm uni}(\bar{\omega})$ [see Fig.\,\ref{fig_dos}\,(c)], and then we obtain
$\bar{T}_c^{\rm VHS} \sim e^{\gamma+\frac{b}{a}-\frac{b}{a}\sqrt{1+\frac{2a}{b^2\bar{U}}}}/\pi$ (see Appendix \ref{sec_a_vhs}).
For $a=0$, $\bar{T}_c^{\rm VHS}$ also reproduces $\bar{T}^{\rm NS}_c$.
For $\bar{U}\to 0$, $\bar{T}^{\rm VHS}_c\propto e^{-\sqrt{2/a\bar{U}}}$, which causes the higher transition temperatures shown in Fig.\,\ref{fig_tn}\,(b) and (c).
The exponential decay for $\bar{U}\to 0$ suggests that the enhancement caused by the VHS is much weaker than that of the FBS (see Appendix \ref{sec_a_universal}).

We further provide the analytical form of $m_{\rm ave}(=\Delta_{\rm ave}/U)$  for any $a$. 
By solving Eq.\,(\ref{eq_gap}) at $T=0$ with $\bar{\rho}_{\rm ave}^{\rm FBS (VHS)}(\bar{\omega})$, we obtain $m_{{\rm ave}, T=0}^{\rm FBS (VHS)}$, written as
$ %\label{eq_mag}
{m}_{{\rm ave}, T=0}^{\rm FBS}=
a\big/
{2b \bar{U} \mathcal{W}
	\big(
	\frac{a}{2b}
	e^{1/b\bar{U}}\big)
}
 $ and $
{m}_{{\rm ave}, T=0}^{\rm VHS}=
e^{\frac{b}{a}-\frac{b}{a}\sqrt{1+\frac{2a}{b^2\bar{U}}-\frac{a^2\pi^2}{6b^2}}}\big/{\bar{U}}$ (see Appendix \ref{sec_a_fbs} and \ref{sec_a_vhs}).
Both $m^{\rm FBS}_{{\rm ave}, T=0}$ and $m^{\rm VHS}_{{\rm ave}, T=0}$ reproduce the non-singular limit $m^{\rm NS}_{{\rm ave}, T=0}=e^{-1/\bar{U}}/\bar{U}$ for $a=0$.
Here, we should note that $m^{\rm FBS}_{\rm ave}\sim a/2+(ab\bar{U}/2)\ln(2/a\bar{U})$ for $\bar{U} \to 0$, and
the constant term $a/2$ explains the specific feature of the flat band magnetism: $m^{\rm FBS}_{{\rm ave}, T=0}$ shows a jump at an infinitesimal $\bar{U}(\sim +0)$ \cite{Noda2014}.

To show the validity of the above qualitative discussions, we quantitatively compare the DMFT calculations with the analytical results.
Figure\,\ref{fig_dos}\,(d) shows the average magnetizations $m_{\rm ave}$ at $T=0$ calculated with the DMFT.
The inset in Fig.\,\ref{fig_dos}\,(d) shows that the analytical forms of $m_{{\rm ave}, T=0}^{\rm NS}$ and $m_{{\rm ave}, T=0}^{\rm VHS}$ with $a=1/3\pi$ agree well with the DMFT calculations.
Later we will discuss the nonmonotonic behavior of magnetization for $\bar{t}_z=0.1$ in Fig.\,\,\ref{fig_dos}\,(d).
Here, we should note that the DMFT does not use the simplified average DOS, suggesting the validity of our simplification with the extraction of the singularity.

Here, we should discuss what determines the singularity weight $a$.
For VHS systems, $a$ can be obtained from the series expansion of $\bar{\rho}_{\rm ave}(\omega)$ at around $\bar{\omega}\sim0$: the present system with $\tilde{t}_z=1$ has $a\sim 1/3\pi$.
For FBS systems, the averaging assumption gives $a$ as follows:
A simple example is the 2D Lieb lattice ($\tilde{t}_z=0$) with $a$ of $1/3$, where one of the three bands is the flat band located at $\bar{\omega}=0$  \footnote{This example can be extended to layered 2D Lieb lattices with an odd number of layers $L=1,3,5, \cdots$, where $a^{}=1/3L$ \cite{Noda2014}.
The present 3D Lieb lattice is the $L\to\infty$ limit of the above, and thus $a=0$. In other words, $a=0$ because all bands are dispersive.}.
On the other hand, for the 3D Lieb lattices ($\tilde{t}_z\not=0$), $a$ is zero because all the bands are dispersive.
However, as shown in Fig.\,\ref{fig_dos}\,(a), there is a very narrow band at around the Fermi energy for small $\tilde{t}_z$.
Within the framework of the static mean-field approximation [Eq.\,(\ref{eq_gap})], the narrow band can be regarded as a flat band when ${U}$ becomes larger than the bandwidth of $4t_z$ (see Appendix \ref{sec_a_eff_fbs}).
To effectively explain such phenomena, we redefine the singularity weight $a$ as a function of the other parameters: $a\to a(\bar{U},\bar{T},\tilde{t}_z)$.

Figure\,\ref{fig_dos}\,(d) shows that, for $\bar{U}\gtrsim 0$, $m_{\rm ave}$ for $\tilde{t}_z=0.1$ rapidly increases from zero without a jump, which can be phenomenologically explained by an increase in $a$ from zero to a finite value as discussed above.
Then, for $\bar{U} \gtrsim 0.1$, $m_{\rm ave}$ increases linearly, which can be effectively explained by the analytical form of $m_{\rm ave}^{\rm FBS}$ with $a\sim0.14$ as shown in the inset of the figure. 
These findings are consistent with $\bar{T}_c$ behavior at around $\bar{U}=0$ and $\bar{T}_c \propto a\bar{U}$ shown in Fig.\,\ref{fig_tn}\,(b).
The behavior of $m_{\rm ave}$ and $\bar{T}_c$ is characteristic of a flat-band magnetism in the 3D layered Lieb lattice \cite{Noda2014}.

Our DMFT calculations in Fig. \ref{fig_tn} (b) and (c) clearly elucidate that, for small $\tilde{t}_z \lesssim 0.1$, $T_c$ is strongly enhanced by the effective FBS resulting from a narrow dispersive band with the interaction effects.
We should stress that this effective FBS can be seen in various systems, such as the multiorbital systems with different bandwidths \cite{Miyahara2007}, and may greatly enhance $T_c$ of magnetism and also superconductivity in these systems.

\begin{figure}[tb]
\includegraphics[clip,width=.6\linewidth]{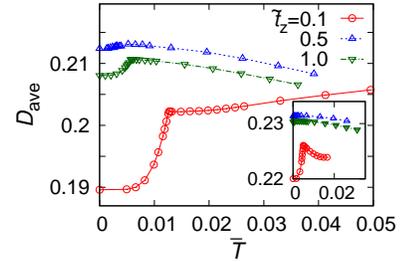}
\caption{(Color online) Average double occupancy $D_{\text{ave}}$ vs. temperature $\bar{T}$ with $\bar{U}=0.2$ for $\tilde{t}_z=0.1,0.5$, and $1.0$. The inset shows that for $\bar{U}=0.1$.}
\label{fig_double}
\end{figure}

We further demonstrate that the introduction of $a(\bar{U},\bar{T},\tilde{t}_z)$ effectively explains the behavior of the thermodynamic quantities.
In Fig.\,\ref{fig_double}, we show the average double occupancy $D_{\text{ave}}=\sum_{\alpha=H,A,B}D_{\alpha}/3$ for $\tilde{t}_z=0.1, 0.5,$ and $1.0$ with $\bar{U}=0.2$.
A kink in $D_{\rm ave}$ clearly shows the transition between magnetic insulating and non-magnetic metallic phases.
At low temperatures, $D_{\rm ave}$ increases with increasing $\bar{T}$;  $\partial D_{\rm ave}/\partial T>0$ for all $\tilde{t}_z$ 
\footnote{In the magnetic phases, $\partial D_{\rm ave}/\partial T$ can be given by $\propto \partial T_c/\partial U$. 
Therefore, $\partial D_{\rm ave}/\partial T >0$ for $\bar{U} \lesssim 1$, and  $\partial D_{\rm ave}/\partial T <0$ for $\bar{U}\gtrsim 1$.}.
At higher temperatures, in the metallic region, we find $\partial D_{\rm ave}/\partial T> 0$ for $\tilde{t}_z=0.1$, whereas $\partial D_{\rm ave}/\partial T< 0$ for $\tilde{t}_z=0.5$ and $1.0$. % ( $\bar{U}=0.2$).
The inset shows that, for smaller $\bar{U}=0.1$, the metallic region shows $\partial D_{\rm ave}/\partial T< 0$ for all $\tilde{t}_z$.
In fact,  $\partial D_{\rm ave}/\partial T< 0$ can be understood from the usual Fermi liquid behavior, while $\partial D_{\rm ave}/\partial T> 0$ is unusual as discussed below.

The thermodynamic relation provides $\partial D/\partial T =-\partial S/\partial U$, where $S$ is entropy.
The Fermi liquid obeys $S\propto M^{\ast} T$, where $M^{\ast}$ is the effective mass.
The interaction-induced mass renormalization means $\partial M^{\ast}/\partial U>0$, which leads to $\partial D_{\rm ave}/\partial T <0$ \cite{Georges1996}.
The FBS breaks down the above scenario.
The entropy is given as $\propto a \ln2+ b M^{\ast} T$, and thus $\partial D_{\rm ave}/\partial T \propto -\partial a/\partial U$ at low temperatures.
We conclude that the unusual behavior $\partial D_{\rm ave}/\partial T> 0$ at $\bar{U}=0.2$ and $\tilde{t}_z=0.1$ results from $\partial a/\partial U <0$, meaning that the renormalization effects greatly reduce the weight of  fermions in the flat (very narrow) band at the Fermi energy.
On the other hand, as discussed above, until ${U}$ becomes comparable to $4{t}_z$, we can expect $\partial a/\partial U >0$, which is consistent with what is shown in the inset of Fig.\,\ref{fig_double}.
Thus, our phenomenological approach explains the unusual $D_{\text{ave}}$ behavior caused by the FBS.
This unusual behavior signals the onset of the specific magnetic transition with $T_c\propto U$.

\begin{figure}[tb]
\includegraphics[clip,width=.98\linewidth]{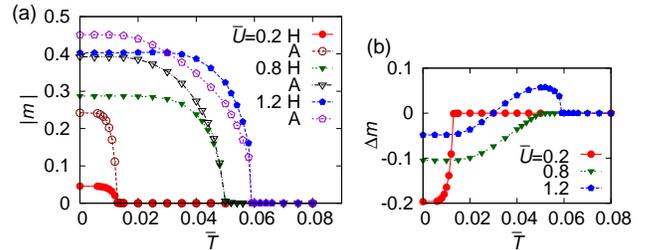}
\caption{(Color online) (a) Magnetization $|m_{\gamma}|$ ($\gamma=H, A$) and (b) the difference of magnetization $\Delta m=|m_H|-|m_A|$ vs. $\bar{T}$ with $\tilde{t}_z=0.1$ for $\bar{U}=0.2,0.8,$ and $1.2$.}
\label{fig_mag}
\end{figure}

We finally discuss $m_{\rm ave}$ near transition temperatures.
For $\bar{U}_c \ll 1$, the gap equation (\ref{eq_gap}) provides
$
m_{{\rm ave}, T\sim T_c}^{\rm FBS}=
({2\sqrt{3}{\bar{T}_c}/\bar{U}})
\big[(\bar{T}_c-\bar{T})/\bar{T}_c \big]^{1/2}
%{\bar{T}_c}
$ except for $a=0$, and
$
m_{{\rm ave}, T\sim T_c}^{\rm VBS}=
({2\sqrt{2}\pi\bar{T}_c}/{\sqrt{7\zeta(3)}\bar{U}})
\big[(\bar{T}_c-\bar{T})/\bar{T}_c \big]^{1/2}
%\sqrt{
%  \frac{\bar{T}_c-\bar{T}}{\bar{T}_c}
%}
$ for any $a$, where $\zeta(x)$ is the Riemann zeta function (see Appendix \ref{sec_a_fbs} and \ref{sec_a_vhs}).
The magnetization is proportional to $(\bar{T}_c-\bar{T})^{1/2}$ even with the singularities, which can be confirmed by the following DMFT calculations.
Figure\,\ref{fig_mag}\,(a) shows $m_{\gamma} (\gamma=H,A)$ for $\bar{U}=0.2, 0.8,$ and $1.2$ with $\tilde{t}_z=0.1$ as a function of $\bar{T}$.

%%%%%%%%%%%%%%%%%%%%%%%%%%
%%%%%%%%%%%%%%%%%%%%%%%%%%%%
%%%%%%%%%%%%%%%%%%%%%%%%%%
%%%%%%%%%%%%%%%%%%%%%%%%%%%%
%%%%%%%%%%%%%%%%%%%%%%%%%%%%%
\section{Strongly interacting region}
\label{sec_stronglyinteracting}
%{\it Strongly interacting region}---
The magnetism in this region is effectively discussed within the local spin picture.
Thus, an important quantity is the site-dependent $\eta_{\alpha}$, namely the number of adjacent sites (i.e., coordination number) in the $xy$ plane of the site $\alpha(=A, B, H)$.
Note that the following discussion will be generally applicable to bipartite lattices with different $\eta_\alpha$.

Employing a simple mean-field approach, we can obtain ${T}_c = {J}_{\rm Hei}/2$ and ${J}_{\rm Hei}= (2\sqrt{\eta_{A} \eta_{H}} t_{xy}^2+ 4t_z^2)/U$, which explains the quantitative change in $\bar{T}_c$ in Fig.\,\ref{fig_tn}\,(b).
Here, $\bar{T}_c$ obeys $\propto \bar{J}_{\rm Hei}=(4\sqrt{2} + 4\tilde{t}_z^2)/(4\sqrt{2}+4\tilde{t}_z)^2(1/\bar{U})$, which takes its minimum value at around $\tilde{t}_z=1.0$.

We also obtain $m_{A (H)}$ near $T_c$ with the mean-field approach: $m_{A (H)}=\pm\sqrt{\frac{3}{2}}\sqrt{\frac{\eta_{A (H)}}{\eta_{A}+\eta_{H}}} \sqrt{\frac{\bar{T}_c-\bar{T}}{\bar{T}_c}}$, meaning that the site with a large $\eta_\alpha$ shows a large $|m_\alpha|$.
Namely, $|m_H|>|m_A|$ for $\bar{U}\gtrsim 1$ at around $\bar{T}_c$, which is confirmed by the DMFT calculations shown in Fig.\,\ref{fig_mag}\,(a).
In contrast, we find $|m_H|<|m_A|$ at low temperatures.
In this region, the quantum fluctuations caused by the itinerancy of electrons yield $\eta_\alpha$-dependent double occupancies $D_H > D_A$, and a large $D_\alpha$ suppresses the development of $|m_\alpha|$ \cite{Note2}.
This causes the crossing curves $m_H$-$\bar{T}$ and $m_A$-$\bar{T}$ seen in Fig.\,\ref{fig_mag}\,(a) for $\bar{U}=1.2$.
Furthermore, we can stress that the change in the sign of $\Delta m=|m_H|-|m_A|$ is a clear manifestation of the crossover of magnetism between band and Heisenberg pictures [see Fig. \ref{fig_mag}\,(b)].

%%%%%%%%%%%%%%%%%%%%
\section{summary}
%%%%%%%%%%%%%%%%%%%%
\label{sec_summary}

We investigated magnetism in three-dimensional layered Lieb lattices and determined the phase diagrams using the dynamical mean-field theory.
We revealed that the (delta-functional) flat-band and (logarithmic) Van Hove singularities affect phase transitions, which greatly increases the transition temperatures for weakly interacting region.
We also pointed out that the effective flat-band singularity emerges from a dispersive flat-band as a consequence of the interaction effects, which can appear in multiband systems.
For strongly interacting region, we characterize $T_c/W$ by the number of adjacent sites.
The larger Heisenberg interaction for site H triggers the onset of magnetization.
Stimulated by this, we proposed a suitable quantity, the difference of magnetization, for clearly detecting the crossover from the flat-band to Heisenberg magnetism, which can be observed in cold atom experiments. 

We proposed a comprehensive approach for investigating the singularity effects by introducing the singularity weight $a$.
We derived the universal forms of $T_c$ and magnetization for both singularities, which offer a remarkable statement: a large singularity-weight induces enhanced $T_c$.
Thus, we elucidated a common feature between the flat-band and Van Hove singularities, which suggests that the singularity weight $a$ is a unified parameter for describing the singularity-induced phase transitions.

\begin{acknowledgments}
We thank N. Kawakami and Y. Takahashi for valuable discussions and R. Peters for his support with the numerical calculations.
This work was supported by JSPS KAKENHI (Grant No. 25287104).
\end{acknowledgments}

%%%%%%%%%%%
\appendix

\section{Derivation of $\bar{T}_c^{\text{FBS}}$}
\label{sec_a_fbs}

\if0
We consider the (rescaled) gap equation
\begin{equation}
\frac{1}{\bar{U}}=\int d\bar{\omega}  \frac{\bar{\rho}_{\rm ave}(\bar{\omega})}{2\sqrt{\bar{\omega}^2+\bar{\Delta}_{\rm ave}^2}}\tanh\bigg(\frac{\sqrt{\bar{\omega}^2+\bar{\Delta}_{\rm ave}^2}}{2\bar{T}}\bigg), \label{eq_gap_re}
\end{equation}
where the notation is the same as in the main text.
\fi

We derive $\bar{T}_c^{\text{FBS}}$ by solving the gap equation Eq.\,(\ref{eq_gap}).
Substituting $\bar{\rho}^{\text{FBS}}(\bar{\omega})= a \delta(\bar{\omega})+ b \theta(1/2-|\bar{\omega}|)$ and $\Delta_{\rm ave}=0$ in Eq.\,(\ref{eq_gap}), we obtain
\begin{equation}
\frac{1}{\bar{U}}
=
\frac{a}{4\bar{T}_{c}}
\int dx \delta(x) \frac{\tanh x}{x}
+ b \int_0^{1/4\bar{T}_c} dx \frac{\tanh x}{x}. \nonumber
\end{equation}
Using $\int_0^{y} dx \tanh x/x = \ln(4e^{\gamma}y/\pi)$ for $y \gtrsim 1$, we obtain $\frac{a\pi}{4b}e^{\frac{1}{b \bar{U}}-\gamma}
=
\frac{a}{4b \bar{T}_c}e^{\frac{a}{4b \bar{T}_c}}, \nonumber
$
where $\gamma$ is the Euler constant.
Then, with the Lambert function ${\cal W}(x)$ defined as $x=\mathcal{W}(x)e^{\mathcal{W}(x)}$, we find $\frac{a}{4b \bar{T}_c}= \mathcal{W}\left(\frac{a\pi}{4b}e^{\frac{1}{b \bar{U}}-\gamma}\right)$.
Finally, we  obtain
\begin{equation}
\label{seq_tc_FBS}
\bar{T}_c^{\rm FBS} =
a \left/ {4b \mathcal{W}\left(\frac{a\pi}{4be^{\gamma}} e^{1/b\bar{U}}\right)}\right.,
\end{equation}
where we assume $\bar{T}_c \lesssim 1$.This form reproduces $\bar{T}_c\propto aU$ for any $a$ and $\bar{T}_c\propto e^{-1/\bar{U}}$ for $a\to0$ as mentioned in Sec. \ref{sec_weaklyinteracting}.

By solving Eq.\,(\ref{eq_gap}) with the same type of calculations as the above,
we can obtain the magnetization $m_{\rm ave}$ at zero temperature:
\begin{equation}
\label{seq_m_FBS}
m_{{\rm ave}, T=0}^{\rm FBS} =
a \left/ {2b\bar{U} \mathcal{W}\left(\frac{a}{2b} e^{1/b\bar{U}}\right)}\right.,
\end{equation}
where we assume $\bar{\Delta}_{\rm ave}=m_{\rm ave} \bar{U} \ll 1$.
It should be noted that $m_{{\rm ave}, T=0}^{\rm FBS}$ shows a finite jump with infinitesimal $U$, which is a distinct feature of the flat-band magnetism.

Employing Eq.\,(\ref{eq_gap}), we can obtain magnetization $m_{\rm ave}$ near the transition temperature:
\begin{equation}
m_{{\rm ave}, T\sim T_c}^{\rm FBS}=
\frac{2\sqrt{3}\pi\bar{T}_c}{\bar{U}}
\sqrt{
  \frac{a+4b^2\bar{T}_c}{a\pi^2+42b\zeta(3)\bar{T}_c}
}
\sqrt{
  \frac{\bar{T}_c-\bar{T}}{\bar{T}_c}
},
\end{equation}
where we use $\bar{\Delta}_{\rm ave}=m_{\rm ave} \bar{U} \ll 1$.
With an additional assumption $\bar{T}_c \ll 1$, we obtain $m_{{\rm ave}, T\sim T_c}^{\rm FBS}\sim
\frac{2\sqrt{3}\bar{T}_c}{\bar{U}}
\sqrt{
  \frac{\bar{T}_c-\bar{T}}{\bar{T}_c}
}
$ except for $a=0$.

%%%%%%%%%%%%%%%%%%%%%%%%%%
\section{Derivation of $\bar{T}_c^{\text{VHS}}$}
\label{sec_a_vhs}

We next derive $\bar{T}_c^{\text{VHS}}$.
Substituting $\bar{\rho}^{\text{VHS}}(\bar{\omega})= \left[a \ln(1/2|\bar{\omega}|)+ b\right] \theta(1/2-|\bar{\omega}|)$ and $\Delta_{\rm ave}=0$ in Eq.\,(\ref{eq_gap}), we obtain
\begin{equation}
\frac{1}{\bar{U}}
=
-a\int_0^{1/4\bar{T}_c} dx  \frac{\ln(4\bar{T_c}x)\tanh x}{x}
+ b \int_0^{1/4\bar{T}_c} dx \frac{\tanh x}{x}. \nonumber
\end{equation}
To solve the above, we give
\begin{eqnarray*}
\int_0^y && \frac{\ln x \tanh x}{x} dx =
 \frac{\ln^2 y}{2}+ A \,\,\,{\rm for}\,\,\, y\gtrsim 1,  \\
A &=& \gamma_1-\frac{\pi^2}{8}-\ln^2\left(\frac{\pi}{2}\right)+\frac{\ln^2 \pi}{2}+\gamma \ln(\frac{\pi}{4}),
\end{eqnarray*}
where $\gamma_1$ is the Stieltjes constant.
Note that we use the following relation to derive the above constant $A$:
\begin{eqnarray*}
\sum_{n=1}^{\infty}(-1)^{n+1} &&\left[\ln^2 n-\ln^2(n+1)\right]=  \\
&&\gamma_1+\frac{\gamma^2}{2}-\frac{\pi^2}{24}+\frac{\ln^2 2 +\ln(4/\pi)\ln \pi}{2}.
\end{eqnarray*}
The gap equation reduces to
$-a A + b \ln(e^\gamma 4/\pi)- [b+a \ln(e^\gamma/\pi)]\ln(4\bar{T}_c)
+\ln^2(4\bar{T}_c)= 1/\bar{U}$.
Then,  we obtain the analytical form
\begin{equation}
\label{seq_tc_VHS}
\bar{T}_c^{\rm VHS}= e^{\gamma+\frac{b}{a}-\frac{b}{a}\sqrt{1+\frac{2a}{b^2\bar{U}}+\frac{a^2C}{b^2}}}/\pi,
\end{equation}
where $C=2A+\ln^2(e^\gamma 4/\pi)=\gamma^2-\pi^2/4+2\ln^2 2+2\gamma_1$.
The above equation holds when $\bar{T}_c \lesssim 1$.

We can obtain the magnetization $m_{\rm ave}$ at zero temperature, by solving Eq.\,(\ref{eq_gap}) similarly:
\begin{equation}
\label{seq_m_VHS}
m_{\rm ave}^{\rm VHS}= e^{\frac{b}{a}-\frac{b}{a}\sqrt{1+\frac{2a}{b^2\bar{U}}-\frac{a^2 \pi^2}{6b^2}}}/\bar{U},
\end{equation}
where we assume $\bar{\Delta}_{\rm ave}=m_{\rm ave} \bar{U} \ll 1$.

Using Eq.\,(\ref{eq_gap}), we can obtain magnetization $m_{\rm ave}$ near the transition temperature:
\begin{eqnarray}
&&m_{{\rm ave}, T\sim T_c}^{\rm VBS}=\frac{2\sqrt{2}\pi\bar{T}_c}{\sqrt{7}\bar{U}} \times
\nonumber
\\
&&
\sqrt{
  \frac{b +a\gamma - a \ln(\pi\bar{T}_c)}{(a+b)\zeta(3)+ a\zeta'(3)- a\zeta(3) \ln(\pi\bar{T}_c/2^{6/7})}
}
\sqrt{
  \frac{\bar{T}_c-\bar{T}}{\bar{T}_c}
},
\nonumber
\\
\end{eqnarray}
where we assume $\bar{\Delta}_{\rm ave}=m_{\rm ave} \bar{U} \ll 1$.
With an additional assumption $\bar{T}_c \ll 1$, we obtain $
m_{{\rm ave}, T\sim T_c}^{\rm VBS}\sim
\frac{2\sqrt{2}\pi\bar{T}_c}{\sqrt{7\zeta(3)}\bar{U}}
\sqrt{
  \frac{\bar{T}_c-\bar{T}}{\bar{T}_c}
}
$ for any $a$.

%%%%%%%%%%%%%%%%%%%%%%%%%%%%%%%%%%%%%%%%%%%%%%%%%%%%%%%%%
\section{Universal ratio}
\label{sec_a_universal}

By using Eqs. (\ref{seq_tc_FBS})-(\ref{seq_m_VHS}), we can discuss the universal ratio $R=2\Delta_{{\rm ave}, T=0}/T_c$, which allows us to quantitatively evaluate the effects of the singularities as discussed below.
It is known that, without the singularity, $R$ stays constant at $2\pi e^{-\gamma}\sim 3.53$, meaning that $\Delta_{T=0}$ and $T_c$ obey a similar form.
Generally speaking, the singularity yields $R$ depending on $U$ and also $a$.

With the FBS, the universal ratio is given by
$$
R^{\rm FBS}=2\pi {\cal W}\left(\frac{a}{2b} e^{1/b\bar{U}} \right)\bigg/{\cal W}\left(\frac{a\pi}{4be^{\gamma}} e^{1/b\bar{U}}\right).
$$
We find that $R^{\rm FBS}$ ranges from $2\pi e^{-\gamma}$ to $4$ depending on $U$ and $a$.
For $\bar{U} \ll 1 $, we obtain $R^{\rm FBS}=2\pi e^{-\gamma}$ $(a=0)$ and $R^{\rm FBS}=4$ $(a\neq0)$.
This jump of $R^{\rm FBS}$ results from the divergent term $e^{1/b\bar{U}}$ in the arguments of Lambert functions.
On the other hand, for the VHS, we obtain
$$
R^{\rm VHS}=2\pi e^{-\gamma} e^{
  \frac{b}{a}\sqrt{1+\frac{2a}{b^2\bar{U}}-\frac{a^2 \pi^2}{6b^2}}
- \frac{b}{a}\sqrt{1+\frac{2a}{b^2\bar{U}}-\frac{a^2 C}{b^2}}
}.
$$
For $\bar{U}\ll1$, we find $R^{\rm VHS}=2\pi e^{-\gamma}$ for any $a$.

These findings indicate that the effects of the FBS (VBS) on the magnetism are strong (weak).
We should note that, for $\bar{U}\ll 1$, the universal ratio depends only on the types of singularities, and does not depend on the weight $a$.
Thus, we can stress that the universal ratio for $\bar{U} \ll 1$ characterizes the strength of the singularity effects.

%%%%%%%%%%%%%%%%%%%%%%%%%%%%%%%%%%%%%%%%%%%%%%%%%%%%%%%%%
\section{Flat band singularity caused by a narrow but finite bandwidth}
\label{sec_a_eff_fbs}

In this section, we explain why the narrow bands can be regarded as the origin of the FBS within the framework of the gap equation. % [Eq.\,(1) in the main text].
We now consider a two-uniform-band (TUB) model, whose
DOS is described as $\bar{\rho}^{\text{TUB}}(\bar{\omega})=(a/\bar{W}_{\rm nar}) \theta(\bar{W}_{\rm nar}/2-|\bar{\omega}|)+b \theta(1/2-|\bar{\omega}|)$, where $\bar{W}_{\rm nar}(=W_{\rm nar}/W < 1)$ is the bandwidth of the narrower band $W_{\rm nar}$ scaled by that of the wider band $W$.
This DOS is naturally regarded as the simplified DOS of the 3D layered Lieb lattice shown in Fig.\ref{fig_dos}\,(a) for $\tilde{t}_z=0.1$ .

As an example, we consider the gap equation for $T_c$, which is given by
\begin{eqnarray*}
\frac{1}{\bar{U}}=\frac{a}{\bar{W}_{\rm nar}}
\int_0^{\bar{W}_{\rm nar}/4\bar{T}_c} dx \frac{\tanh x}{x}
+ b \int_0^{1/4\bar{T}_c} dx \frac{\tanh x}{x}.
\end{eqnarray*}
When both $\bar{W}_{\rm nar}  \gtrsim  4\bar{T}_c$ and $1  \gtrsim  4\bar{T}_c$ are satisfied, given $\int_0^{y} dx \tanh x/x = \ln(4e^{\gamma}y/\pi)$ for $y \gtrsim 1$, the gap equation produces:
$$
\bar{T}_c^{\rm TUB}=\frac{e^\gamma}{\pi} \bar{W}_{\rm nar}^\frac{a}{a+b \bar{W}_{\rm nar}} e^{-\frac{\bar{W}_{\rm nar}}{ \bar{U}(a+b \bar{W}_{\rm nar})}}.$$
This form is qualitatively equivalent to the non-singular form $\bar{T}_c^{\rm NS} \propto e^{-1/\bar{\rho}(E_F)\bar{U}}$, where $\bar{\rho}(E_F)=(a+b \bar{W}_{\rm nar})/\bar{W}_{\rm nar}$.

If $\bar{W}_{\rm nar} \lesssim 4\bar{T}_c$ and $1 \gtrsim 4\bar{T}_c$ are satisfied,
by using another relation $\int_0^{y} dx \tanh x/x = y$ for $y \lesssim 1$, we obtain $\frac{a\pi}{4b}e^{\frac{1}{b \bar{U}}-\gamma}
=
\frac{a}{4b \bar{T}_c}e^{\frac{a}{4b \bar{T}_c}}
$.
This gap equation is the same as that for the FBS as shown above in Appendix \ref{sec_a_fbs}, which leads to $T_c^{\rm FBS} \sim  aU/4$.
%\begin{eqnarray*}
%\int_0^{y} dx \frac{\tanh x}{x} &=& y \,\,\, {\rm for}\,\,\, y \lesssim 1 \\
%&=& \ln(4e^\gamma y/\pi) \,\,\, {\rm for}\,\,\, y \gtrsim 1.
%\end{eqnarray*}
Thus, when the interaction strength becomes greater than the narrower bandwidth, $\bar{U} \gtrsim \bar{W}_{\rm nar}$, we find the linear-$U$ behavior of $T_c$.
We should note that here we use a relation $\bar{U} \propto 4\bar{T}_c$.
Consequently, we can conclude that the very narrow band can be regarded as the origin of the (remaining) FBS as mentioned in Sec. \ref{sec_weaklyinteracting}.

Here, we should comment that the above discussion is based on the static mean-field approximation.
Thus, we miss some effects caused by the interactions, such as the mass renormalization and the creation of the Hubbard bands.
Simply put, the above gap equation cannot be used for the strongly interacting (Heisenberg) region, and thus, this approach may overestimate the region in which $T_c\propto U$ behavior can be obtained.
Note that the DMFT calculations properly capture the effects mentioned above, and the DMFT calculations confirm that $T_c\propto U$ behavior is obtained for small $\tilde{t}_z (\lesssim 0.1)$.

By extending the simple gap equation, we can discuss the above phenomena effectively.
Here, we redefine $a$ as a function of other parameters: $a\to a(\bar{U},\bar{T},\tilde{t}_z)$.
For example, we can expect that $a$ the quasi-particle weight of the fermions in the narrow band will decrease owing to the renormalization effects.
This phenomenological approach can explain the characteristic behavior of thermodynamic quantities as shown in Fig.\,\ref{fig_double} in the main text.

%\bibliography{3dlieb.bib}

\end{document}